\documentstyle[prl,aps,psfig]{revtex}

\begin{document}

\draft

\title{Acoustic collective excitations and static dielectric response\\
in incommensurate crystals with real order parameter}

\author{V. Danani\'{c}}  
 \address{ Department of Physics, Faculty of Chemical Engineering and 
 Technology,  University of Zagreb\\
Maruli\'{c}ev trg 19, 10000 Zagreb, Croatia}
\author{A. Bjeli\v{s} and M. Latkovi\'{c}}
 \address{Department of Theoretical Physics, Faculty of Science, University 
 of Zagreb\\
 Bijeni\v{c}ka 32, 10000 Zagreb, Croatia}

\maketitle

\bigskip
\bigskip

\centerline{\bf Dedicated to Professor Boran Leonti\'{c} on the occasion of
his 70$^{\rm th}$ birthday}

\begin{abstract}

Starting from the basic Landau model for the incommensurate-commensurate 
materials of the class II, we derive the spectrum of collective modes for all 
(meta)stable states from the corresponding phase diagram. It is shown that 
all incommensurate states posses Goldstone modes with acoustic dispersions. 
The representation in terms of collective modes is also used in the calculation 
and discussion of static dielectric response for systems with the commensurate 
wave number in the center of the Brillouin zone. 

\end{abstract}

\bigskip


\section{Introduction}

The uniaxial incommensurate-commensurate (IC) materials of the II class 
have the wave number of commensurate ordering,  $q_c$, either at the center 
($q_c = 0$) or at the border ($q_c = \pi$) of the Brillouin zone.  Here the 
unit length is taken equal to the lattice constant. The corresponding Landau 
free energy functional is then an expansion in terms of a real order parameter 
$u(z)$~\cite{hornreich,michelson,is1,dbcr},
\begin{equation}
 f[u]=\frac{1}{2 L}\int_{- L}^{L}
 \left[c\left(\frac{d u}{d z}\right)^{2} + \,  
 d\left(\frac{d^{2}u}{dz^{2}}\right)^{2} +
\, a u^{2} + \,{\frac{1}{2}} b u^{4}\right]dz\,\, .
\label{fe}
\end{equation}
Since the parameter $c$ may acquire negative values, it is necessary to 
include into the expansion~(\ref{fe}) the further, second derivative term, 
with a coefficient $d$ presumably positive.

The mean-field phase diagram follows from the simple variational 
procedure for the functional~(\ref{fe}). It contains disordered, commensurate, 
and various stable~\cite{michelson,is1,bcm} and metastable~\cite{dbcr}
periodic phases. The most important among the latter is almost sinusoidal, 
\begin{equation}
u_s(z) \approx \sqrt{\frac{4}{3b}\left(\frac{c^{2}}{4d}-a\right)}
     \sin\left(\sqrt{-\frac{c}{2d}}z\right)\, ,
\label{us}
\end{equation}
with weak higher harmonics~\cite{michelson,dbcr}.

Although represented in terms of the one-component (real) order parameter, 
the periodic phases like~(\ref{us}) are generally incommensurate with respect 
to the underlying lattice. From the other side, it was stated~\cite{adp}
that incommensurate orderings, 
including those close to the commensurabilities of the II class, should 
have to be represented by at least two quantities, i.e., by
the amplitude and the phase of some periodic modulation. This expectation 
originates from the experience with the I class of IC systems, characterized
by at least two-dimensional order parameters. More specifically, the 
incommensurate orderings are then most often represented by the modulation of 
the phase of a complex order parameter. 
Since the model~(\ref{fe}) apparently does not comprise a phase variable, it 
was interpreted as a reduction, appropriate only for the description of the 
commensurate ordering, of some richer physical models that include at least 
two coupled modes, i.e., one-component order parameters. The explicit
Landau model with this property was formulated by Levanyuk and 
Sannikov~\cite{ls}, and widely explored afterwards~\cite{tol,adp,mash,ss}.
The alternative approaches along same lines, attempting primarily to explain 
the phase diagram for betaine calcium chloride dihydrate (BCCD), were also 
proposed~\cite{rib,mor}. The physical arguments for such general approach to 
insulators from the II class were given by Heine and McConnell~\cite{hmc}. 

Coming back to the model~(\ref{fe}), it is important to emphasize that,
either already bearing the whole physical relevance (like in charge
density wave materials~\cite{braz}), or being derived from a more complex
starting scheme~\cite{hmc}, it accounts for the cross-over between
orderings with real and complex order parameters. In that sense the
solutions like~(\ref{us}) are  
examples of the mean field approximation for the latter. This can be most 
easily recognized by looking at the dispersion of the quadratic 
part of the expansion~(\ref{fe}) in the reciprocal space, with the 
biquadratic ("bottle bottom") dependence on the wave number (see 
Eq.~\ref{dis}). This dispersion is an expansion around a pair of symmetry 
related minima, that in addition takes properly into account the symmetry 
condition on evenness with respect to the center (or the boundary) of the 
Brillouin zone. In this sense the model~(\ref{fe}) can be qualified as the 
basic one for the II class, while the more complex Landau expansions~\cite{ls} 
are necessary in, physically possible, but non-generic cases when 
two or more hybridized modes are simultaneously close to instability.

In order to justify the above statement at the mean field level, one has to 
prove that incommensurate periodic states of the model~(\ref{fe}) fulfill 
the crucial general property of incommensurately modulated orderings,
namely that for each of them there exists a Goldstone mode with the frequency 
$\Omega_0 (k)$ that vanishes at $k = 0$, and generally has a finite slope 
(phase velocity) $\partial \Omega_0 (k)/\partial k \equiv v_G $ in the long 
wavelength limit $k \rightarrow 0$. Such mode should exist in the whole control 
parameter (e.g. temperature) range of (meta)stability for a given state. It 
would correspond to standard acoustic phason branches for IC orderings in 
the materials of the I class. The existence of the Goldstone mode in 
incommensurate states like~(\ref{us}) for the materials of the II class
is to be contrasted to a widely accepted belief~\cite{mash} that these states 
do not have a phason mode. 

The requirement $\Omega_0 (k=0) = 0$ guarantees the presence of translational 
degeneracy of the ordered state. In the particular case~(\ref{fe}) this 
means that the periodic solutions should be invariant to translations 
$z \rightarrow z + z_0$ with arbitrary $z_0$. This is obviously fulfilled, 
since the kernel in the functional~(\ref{fe}) does not depend explicitly 
on $z$. 

It remains to find out whether there exists a Goldstone mode with a 
{\em nontrivial} dispersion [$\Omega_0 (k) \neq 0$ for $k\neq 0$], and, if 
so, to determine its properties at the critical lines of the phase diagram for 
the model~(\ref{fe}). With this aim, we calculate in the present work the 
spectra of collective modes for all basic orderings that follow from the 
corresponding Euler-Lagrange equation. To this end we generalize the
usual eigenvalue
problem for collective modes to the systems with non-standard free energy
densities, in particular to those like~(\ref{fe}) with higher order terms in 
the gradient expansion. Details of this method are presented in Ref.~\cite{dbl}.
Combining analytic and numerical analysis we show 
that the Goldstone modes for the incommensurate orderings like~(\ref{us}) 
exist in the whole range of their stability. Furthermore, it appears
that collective modes of the model~(\ref{fe}) have some peculiar, 
experimentally observable, properties. For illustration we discuss 
the contribution of collective modes to the dielectric response of the 
II class materials for which the order parameter coincides with their 
electric polarization. 

In Sec.~II we continue by considering collective modes for the sinusoidal 
state~(\ref{us}) and for the disordered state of the free energy~(\ref{fe}). 
The expressions for corresponding dielectric susceptibilities are derived in 
Sec.~III. The concluding remarks are given in Sec.~IV.

\section{Collective modes}

The Euler-Lagrange equation for the model~(\ref{fe}) reads  
\begin{equation}
d\frac{d^{4}u}{dz^{4}}+\,c\frac{d^{2}u}{dz^{2}}+ \,a u+ \,b u^{3}=0.
\label{el}
\end{equation}
Given a solution $u_0(z)$ of this equation that also satisfies additional
extremalization requirements~\cite{db}, the corresponding eigenvalue 
problem is defined by 
\begin{equation} 
{\cal D}\eta \equiv d\eta''''_{\Omega}(z)
-c\eta''_{\Omega}(z)+\left[a+3bu_{0}^{2}(z)\right]\eta_{\Omega}(z) = 
\Omega^2\eta_{\Omega}(z).
\label{eta}
\end{equation}
The spectrum of collective modes for a given (meta)stable solution $u_0(z)$ is
defined by those non-negative values of $\Omega^2$ for which the
problem~(\ref{eta}) has normalizable solutions $\eta_{\Omega}(z)$. For periodic 
orderings, $u_0(z + 2\pi/Q) = u_0(z)$, we adopt the Floquet analysis of this 
problem~\cite{dbl}. In particular, it is convenient to use for such orderings 
the Bloch representation $\eta_{\Omega}(z) \rightarrow \eta_{n,k}(z)$, with
\begin{equation}
\eta_{n,k}(z)=e^{ikz}\Psi_{n,k}(z),\,\,\,\,\,\,\,\,
\Psi_{n,k}\left(z+\frac{2\pi}{Q}\right)=\Psi_{n,k}(z).
\label{bloch}
\end{equation}
The wave number $k$ is limited to the first Brillouin zone formed by the
incommensurate modulation, $-Q/2\leq k \leq Q/2$,
and $n$ is the branch index. The formulation~(\ref{bloch}) enables the 
extension of the numerical method, developed for finding the solutions 
$u_0(z)$~\cite{dbcr}, to the calculation of dispersions $\Omega_{n}(k)$ 
and Bloch functions $\Psi_{n,k}(z)$. 

While the previous calculations of the spectrum of collective modes for the 
incommensurate solution $u_s$ given by Eq.~\ref{us} were 
approximative~\cite{ista}, the present approach~(\ref{eta},\ref{bloch}) enables 
exact results, as shown in  Fig.~1. The necessary condition for the 
(meta)stability of this and other periodic solutions is $c < 0$. Then it 
is convenient to introduce the parameter $\lambda \equiv ad/c^2$ and reduce 
the model~(\ref{fe}) to a single parameter problem~\cite{dbcr}. The lowest 
among branches from Fig.~1, that with the long wavelength dispersion 
$\Omega_0 (k) = v_G k$, is the Goldstone mode. As is seen from 
Fig.~1c, the phase velocity $v_G$ tends to zero as $\lambda$ approaches the 
lower stability edge at $\lambda_{s} = -1.835$ (Fig.~1a), and remains finite 
at the second order phase transition to the disordered state at 
$\lambda_{id} = 0.25$ (Fig.~1b). In the latter figure we use both, Brillouin 
and extended, schemes for the $k$-space, since the former becomes irrelevant 
for $\lambda \geq \lambda_{id}$. The collective modes then reduce to the 
unique dispersion curve for the disordered state, 
\begin{equation}
\Omega_{d}^{2}(k) = a + c k^2 + d k^4 , 
\label{dis}
\end{equation}
defined by the gradient expansion in Eq.~\ref{fe}. 

Other collective modes for $u_s$ are massive. The dispersion curves for three
of them are also shown in Fig.~1. The gap of the lowest lying one, 
$\Omega_{1}(0)$, tends to zero as $\lambda \rightarrow \lambda_{id}$. As is seen
in Fig.~1b, at the very transition to the disordered state, 
$\lambda = \lambda_{id}$, this mode has an acoustic dispersion, with the phase 
velocity equal to $v_G$, i.e., to that of the Goldstone mode.

The spectra of collective modes for other, metastable, periodic solutions 
of Eq.~\ref{el} resemble to that of $u_s$. In particular, for all of them the
phase velocities of Goldstone modes vanish as the parameter $\lambda$ tends 
towards both edges of metastability for a given solution~\cite{dbl}. This 
is to be contrasted to $v_G$ for $u_s$, which remains finite at one edge, 
i.e., at $\lambda_{id}$ (Fig.~1c).

\section{Dielectric susceptibility}
 
The straightforward use of the above Bloch basis~(\ref{bloch}) leads to the 
dielectric susceptibility for the incommensurate state $u_s$, expressed in 
terms of branches of collective modes. In the static limit which 
is under consideration here, it is given by 
\begin{equation} 
\alpha_i\,\,\, = \,\,\,\frac{d}{c^{2}}\left[\frac{|g|^2}{v_{G}^{2}} + 
\sum_{n \neq 0} \frac{|\psi_{n}|^2}{\Omega_{n}^{2}(0)}\right]\,\,\,.
\label{sci}
\end{equation}
The first term is the response from the Goldstone mode. Here
$g \equiv \frac{1}{2L}\int_{-L}^{L} \Psi_{0}^{(1)}(z)dz$, and 
$\Psi_{0}^{(1)}(z)$ is the coefficient in the long wavelength expansion of 
the corresponding Bloch function~(\ref{bloch}), 
\begin{equation}
\Psi_{0,k}(z) \,\,= \frac{1}{\sqrt{{\cal N}}}\,\,\frac{d u_s (z)}{d z}\,\,\ 
+\,\, k\Psi_{0}^{(1)}(z)+... 
\label{exp}
\end{equation}
where ${\cal N}$ is the normalization constant. The expansion of ${\cal N}$ in 
powers of $k$ does not contain the term linear in $k$, i.e., in the lowest 
order approximation it is given by
\begin{equation}
{\cal N} =\frac{1}{2L}\int_{-L}^{L}\left(\frac{du_s(z)}{dz}\right)^2dz\,\,.
\label{norm}
\end{equation}
The residual sum in Eq.~\ref{sci} includes the contributions from massive 
collective modes, i.e., those with finite gaps $\Omega_{n}(0)$. 
Corresponding oscillatory strengths are given by
\begin{equation}
|\psi_{n}|^2 \,\,\,\equiv \,\,\,\arrowvert\frac{1}{2L}
\int_{-L}^{L}\Psi_{n,k = 0}(z)dz\arrowvert^{2} \, .
\label{ost}
\end{equation}

It can be shown~\cite{dbl1} that the coefficient $g$ remains finite as 
$\lambda$ approaches the stability edge at $\lambda_s =-1.835$, so that then 
$\alpha_i$ diverges as $v_{G}^{-2}$. From the other side, this coefficient 
vanishes together with the function $\Psi_{1}(z)$ at the transition to 
the disordered state ($\lambda = \lambda_{id} = 0.25$). Since $v_G$ remains 
constant, it becomes clear from the expression~(\ref{sci}) that in this limit 
the finite contribution to $\alpha_i$ comes from the residual sum. When 
calculated directly, by expanding the linear response equation in powers of 
$\lambda_{id} - \lambda$, this contribution reads~\cite{dbl1}
\begin{equation} 
\alpha _{i}(\lambda) \approx \frac{d}{c^{2}}\,\,\frac{1}{\frac{1}{2}-\lambda}
\left[1 + \frac{2\left(\lambda_{id} - \lambda\right)^{2}}{\left(\frac{1}{2}
 - \lambda\right)\left(\frac{5}{2}-\lambda\right)}+...\right]\,\,.\,\,
\label{scp}
\end{equation}
The previous calculations~\cite{is1,ioi} were limited only to the leading
term in this expansion.

The careful inspection of the 
residual sum in~(\ref{sci}) shows that it contributes 
to the susceptibility~(\ref{scp}) only via one mode, denoted by $\Omega_2$ in
Fig.~1. We note that the oscillatory strength~(\ref{ost}) of the lowest massive 
mode $\Omega_1$ vanishes, so that it does not contribute to the
susceptibility~(\ref{sci}). The mode  
$\Omega_2$ "survives" the phase transition at $\lambda_{id}$,
and remains the only contribution, presented by the dashed curve in
Fig.~1b. The dielectric susceptibility in the disordered state at 
$\lambda > \lambda_{id}$ is given by 
\begin{equation}
\alpha_d = \frac{1}{\Omega_{d}^{2}(0)} = 
\frac{1}{a} \,\,\,\,.
\label{scd}
\end{equation}
Thus, at the phase transition from the incommensurate to the 
disordered state the dielectric susceptibility varies continuously, but with 
a finite jump in $\frac{d \alpha}{d \lambda}$ at 
$\lambda = \lambda_{id}$. 

To summarize the above discussion of the dielectric responses
$\alpha_i$ and $\alpha_d$, we show in Fig.~2 their dependence on the 
parameter $a$ (i.e. on temperature). Other parameters are fixed.
Fig.~2 also includes the susceptibility for the commensurate solution 
$u_c = \pm \sqrt{-a/b}$, which is stable in the ranges $\lambda < -\frac{1}{8}$ 
for $c < 0$, and $a < 0$ for $c > 0$. It is separated from the disordered 
state $u_d$ by the second order transition at the line $a = 0, \,\, c > 0$, 
and from the incommensurate state $u_s$ by the first order transition, defined 
by the line $c < 0, \,\,\, \lambda \approx -1.177$.

The static susceptibility of the commensurate state $u_c$ is given by
\begin{equation}
\alpha_{c} =  \frac{1}{a+3bu_{c}^{2}} = \,\,\frac{1}{\Omega_{c}^{2}(0)} = 
\,\,-\,\,\frac{1}{2a}\,\,, 
\label{scc}
\end{equation}
where 
\begin{equation}
\Omega_{c}^{2}(k) = d k^4 + c k^2 - 2 a 
\label{com}
\end{equation}
is the dispersion for the corresponding unique collective mode that follows 
from Eq.~\ref{eta} after inserting $u_0 (z) = u_c$. We see from
Eqs.~(\ref{scd}) and~(\ref{scc}) that the static susceptibility shows a 
standard type of divergence at the line of the second order transition from the 
disordered to the commensurate state, $a = 0, \,\, c > 0$. From the other side, 
$\alpha_{c}$ has a finite value at the stability edge ($c < 0, 
\lambda_c = -\frac{1}{8}$) for $u_c$ in the regime of coexistence with the 
incommensurate state $u_s$. Thus, we propose an asymmetric behavior of the 
susceptibility as one passes through the hysteresis range (in, e.g.,
temperature), bounded by the values $\lambda_c$ and $\lambda_s$ in the 
parameter $\lambda$. Namely, as shown in Fig.~2, by cooling through the 
incommensurate state one ends with the divergence of $\alpha_i$ before passing 
into the commensurate state at $\lambda_s$, while by heating through the 
commensurate state one passes into the incommensurate state at $\lambda_c$ with 
the jump in the susceptibility equal to 
$[\alpha_c - \alpha_i]_{\lambda = \lambda_c}$.

Due to the existence of other metastable periodic solutions~\cite{dbcr} in 
the above coexistence range of the parameter $\lambda$, the temperature 
variation  of the susceptibility may be even more complicated than that 
schematically presented in Fig.~2. The more detailed analysis~\cite{dbl} 
shows that, provided that for a given material and in particular
circumstances some of these states are stabilized, the static susceptibility 
will diverge at both edges of their stability ranges, again, like in the case 
of the state $u_s$, due to the vanishing of phase velocity for the 
corresponding Goldstone modes. Since these stability ranges are rather narrow 
in comparison with $\lambda_c - \lambda_s \approx 1.71$, the stabilization of 
highly nonsinusoidal periodic states~\cite{dbcr} would be manifested by 
dramatic variation of susceptibility in relatively short temperature intervals.

\section{Conclusion}

In the above analysis it was clearly shown that the model~(\ref{fe}), 
although formulated in terms of a single and  real order parameter, provides 
the existence of the acoustic Goldstone mode for the incommensurate ordering. 
For small wave
numbers ($k \ll Q$) this mode simply generates compressions and dilatations 
of the sinusoidal modulation $u_s (z)$, i.e., its physical content is same as
that of standard phason mode for incommensurate orderings in the II class 
of IC systems. Amplitude fluctuations of this modulation are generated  through 
higher, massive modes from Fig.~1. Still, the lowest among these massive modes 
have some peculiar properties. For one of them the gap tends to zero by 
approaching the transition from, e.g., the sinusoidal state to the disordered 
state. However this mode is not optically active. The whole optical activity
at this transition comes from the another massive mode in order, as shown in 
Fig.~1.

In contrast to standard approaches~\cite{blinc}, the dielectric functions 
for all (meta)stable states are here strictly represented in terms of 
collective modes. We are thus able to distinguish modes with finite
contributions in the optical response from those with no dipolar polarization.
Note that the present analysis of dielectric response is limited to systems 
with uniform commensurate orderings (those with $q_c = 0$). The extension to 
systems with dimerizations, as well as the equivalent treatment of collective 
modes and dielectric response of extended Landau models for IC systems of 
class I~\cite{lb}, are under current investigations.  

The above analysis shows that incommensurate states that follow already from 
the simplest basic version of the model~(\ref{fe}) for the IC systems of class 
II, have the same physical properties as those of class I, represented in 
terms of the complex order parameter. This model is still insufficient for the
explanation of rich phase diagrams of some well-known representatives of class 
II, like thiourea and BCCD. In this respect the question which remains is, how
to make an appropriate extension without invoking an additional order
parameter, in a way analogous to the recent proposal for the class I~\cite{lb}, 
that would as well stabilize other commensurate states with wave numbers 
close to $q_c = 0$  or $q_c = \pi/a$. 

\bigskip

{\bf ACKNOWLEDGMENTS}

The work was supported by the Ministry of Science and Technology of 
the Republic of Croatia through project No. 119201.

\newpage

\begin{figure}
\caption{ The dispersion curves for the lowest collective modes of the 
incommensurate state~(\ref{us}), for $\lambda= \lambda_s= -\,1.8$ (full lines), 
$\lambda = -\,0.5$ (dashed lines) (Fig.~1a), and for
$\lambda=\lambda_{id} = \,0.25$ in the Brillouin (full lines) and extended 
(dashed line) schemes (Fig.~1b).  The dependence of the phase velocity of the 
Goldstone mode on the parameter $\lambda$ is shown in Fig.~1c.}
\bigskip
\centerline{\psfig{file=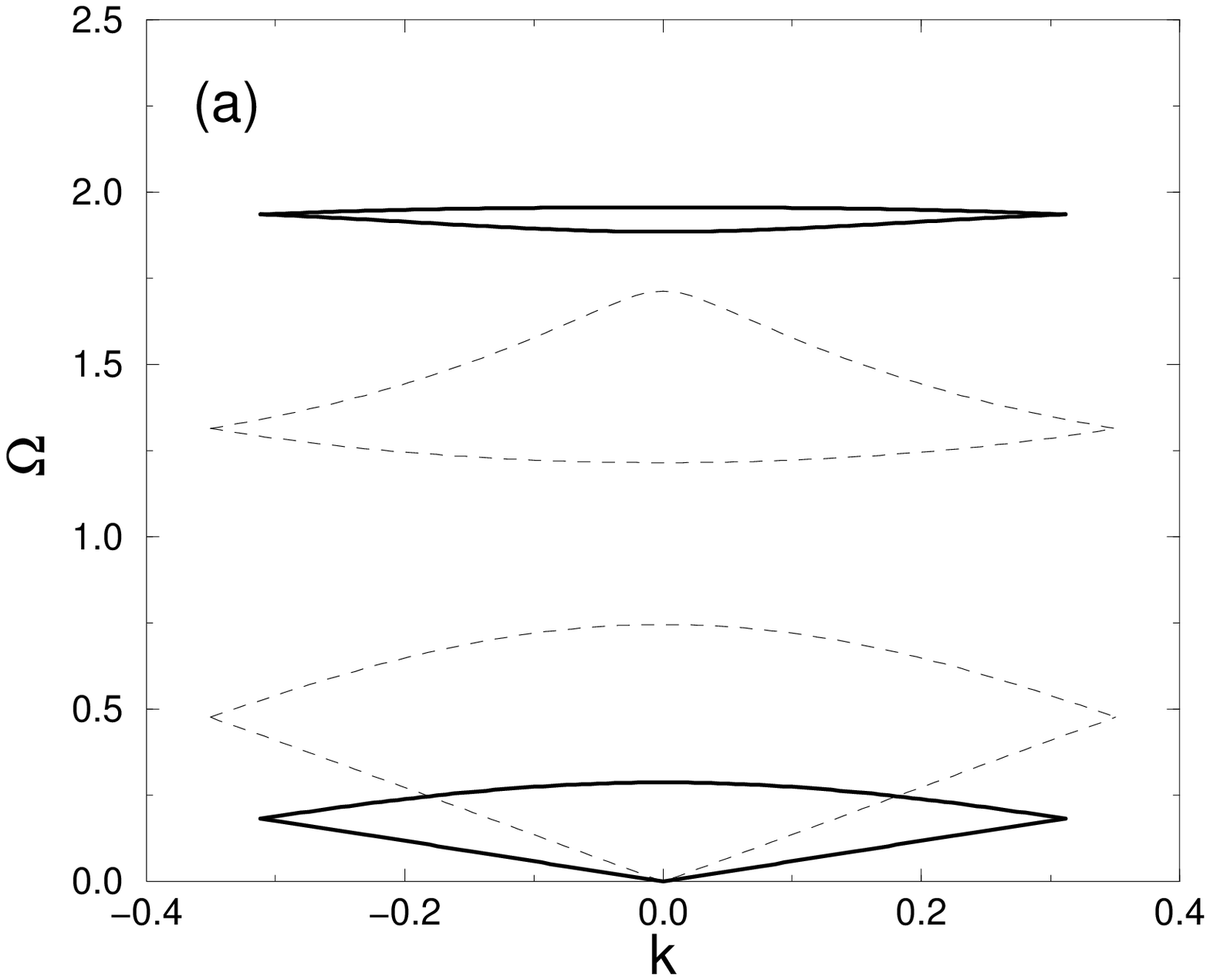,height=10cm,width=12cm,silent=}}
\centerline{\psfig{file=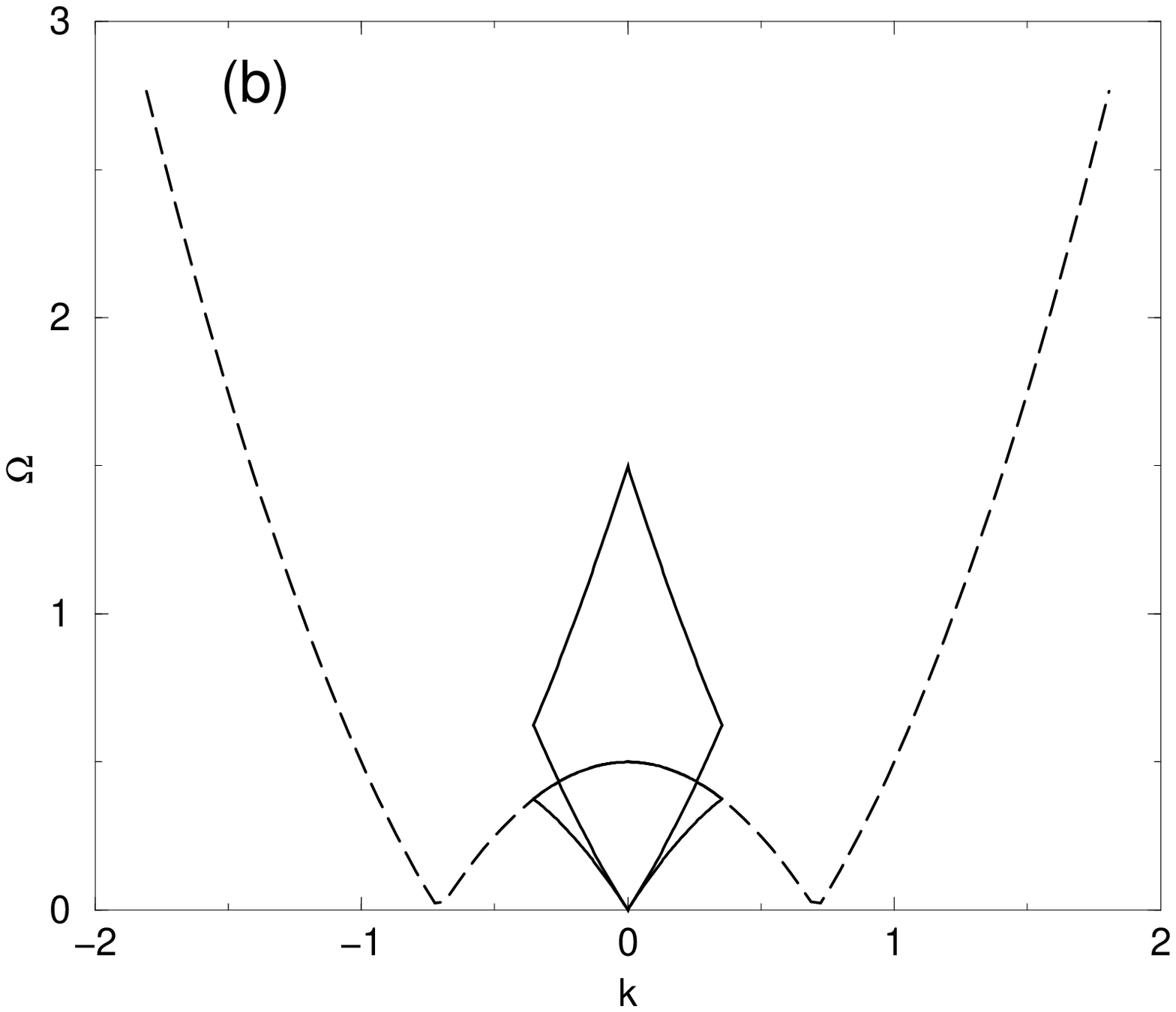,height=12cm,width=14cm,silent=}}
\centerline{\psfig{file=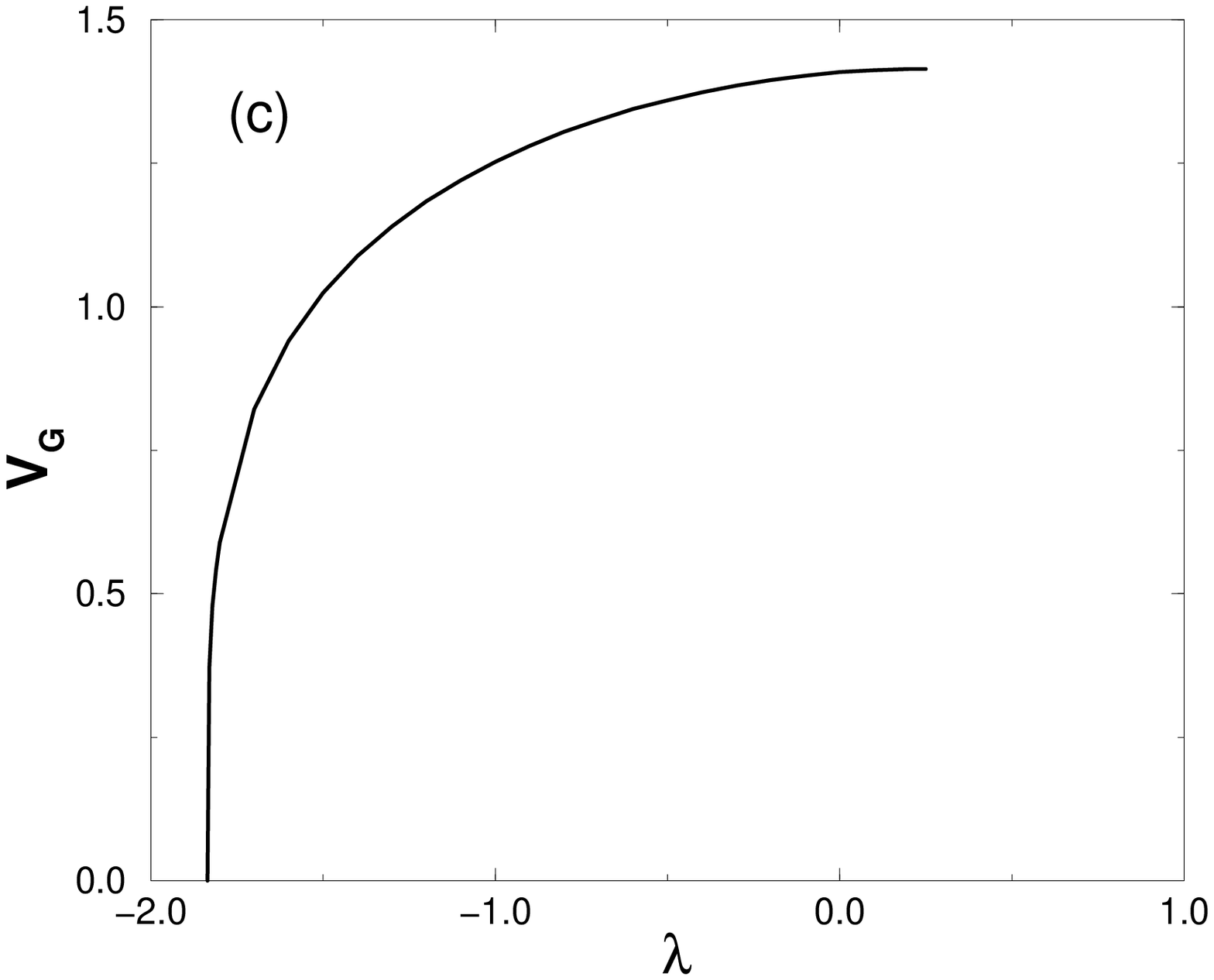,height=10cm,width=12cm,silent=}}
\label{fig1}
\end{figure}

\begin{figure}
\caption{ Dielectric susceptibilities for the incommensurate state (full line), 
commensurate state (dashed line), and disordered state (dotted line) as 
functions of the parameter $\lambda$. The critical values of $\lambda$ shown in
the figure, are introduced in the text.}
\bigskip
\bigskip
\centerline{\psfig{file=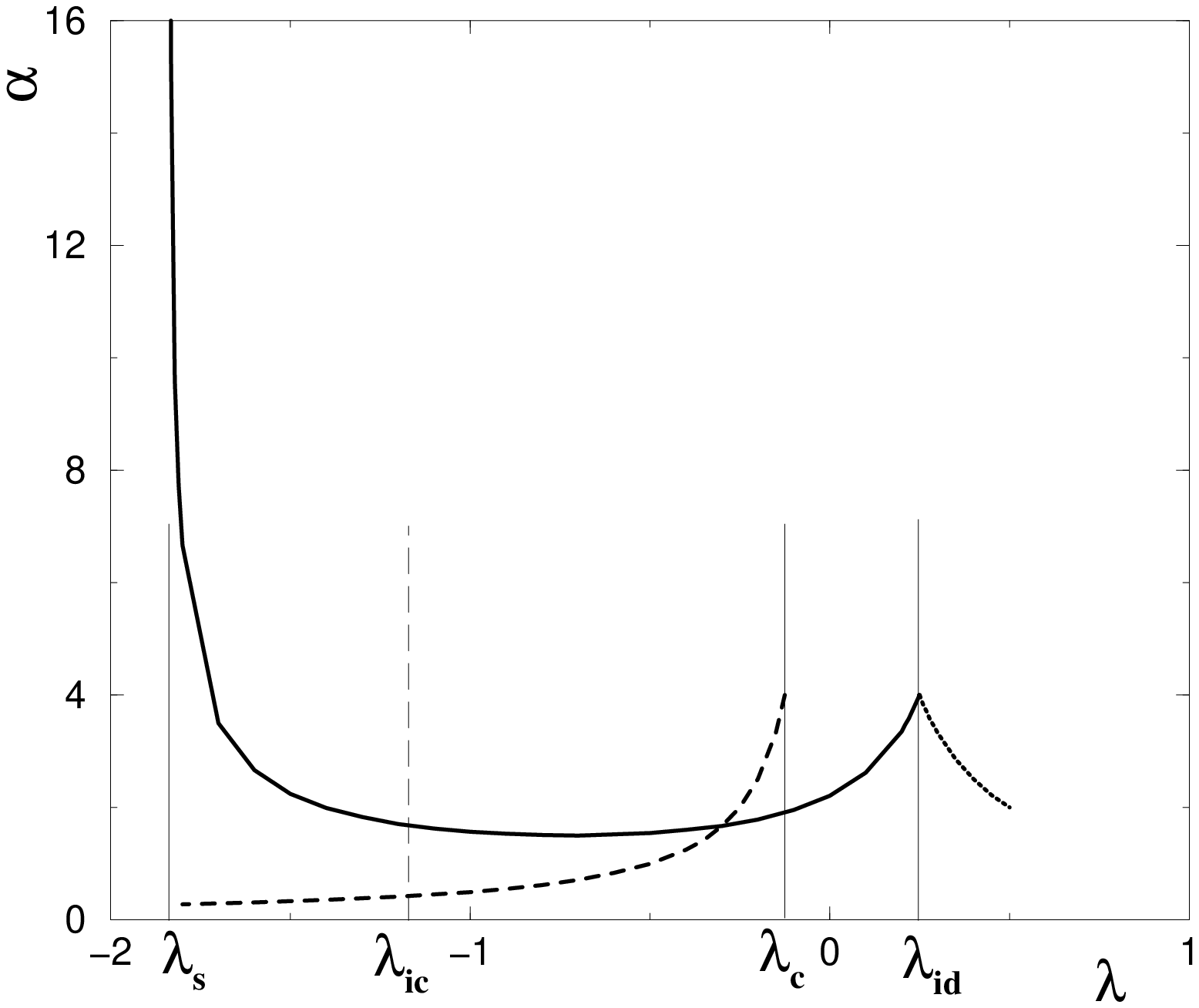,height=10cm,width=12cm,silent=}}
\label{fig2}
\end{figure}

\end{document}